\documentclass[twocolumn,amsmath,amssymb,prl,10pt,nofootinbib,superscriptaddress]{revtex4}
\usepackage{ graphicx, float,amsmath, amsmath, amssymb}
\usepackage[export]{adjustbox}

\def\be{\begin{equation}}
\def\ee{\end{equation}}
\def\bea{\begin{eqnarray}}
\def\eea{\end{eqnarray}}
\def\bse{\begin{subequations}}
\def\ese{\end{subequations}}

\usepackage[breaklinks, colorlinks, citecolor=blue]{hyperref}

\usepackage[labelsep=period]{caption}
\usepackage[normalem]{ulem}
\usepackage{soul}
\usepackage{minitoc}

\usepackage{bm}

\begin{document}
\title{A pure general relativistic non-singular bouncing origin for the Universe}

\author{Aur\'elien Barrau}%
\affiliation{%
Laboratoire de Physique Subatomique et de Cosmologie, Universit\'e Grenoble-Alpes, CNRS/IN2P3\\
53, avenue des Martyrs, 38026 Grenoble cedex, France}

%\author{Killian Martineau}%
%\affiliation{
%Laboratoire de Physique Subatomique et de Cosmologie, Universit\'e Grenoble-Alpes, CNRS/IN2P3\\
%53, avenue des Martyrs, 38026 Grenoble cedex, France
%}

%\author{Selim Touati}%
%\affiliation{%
%Laboratoire de Physique Subatomique et de Cosmologie, Universit\'e Grenoble-Alpes, CNRS/IN2P3\\
%53, avenue des Martyrs, 38026 Grenoble cedex, France
%}

\date{\today}
\begin{abstract} 
In this article, we argue that the past of the Universe, extrapolated from standard physics and measured cosmological parameters, might be a non-singular bounce. We also show that, in this framework, quite stringent constraints can be put on the reheating temperature and number of inflationary e-folds, basically fixing $T_{RH}\sim T_{GUT}$ and $N\sim 70$. We draw some conclusions about the shape of the inflaton potential and raise the ``naturalness" issue in this context. Finally, we argue that this could open a very specific window on the ``pre big bounce" universe.
\end{abstract}
\maketitle

\section{Introduction}

Cosmological studies often mix the search for the correct theory required to described the World and the investigation of contingent phenomena occurring within a well defined paradigm. In many cases, it is indeed welcome and legitimate not to distinguish between both questions that are somehow interconnected. In this article, however, we make a clear choice: we assume the laws of physics to be known and described by the standard models. We also take for granted the usual ingredients of the cosmic scenario, in particular the existence of an inflationary stage. As a less consensual ingredient, although now supported by observations, we finally add a positive curvature to the Friedmann equation. Although still debated, the closedness of the Universe is quite strongly favored by recent data and new analyses, as explained in the next section.\\

The point of view adopted here is therefore the following: we simply try to trace back the history of the Universe to understand what has happened close to the Big Bang. We  work in the very same way than a paleontologist or a historian would, trying to determine the correct past without focusing on figuring out if the specific history observed can be considered as probable or not. Trying to understand why things are what they are is, in principle, a very different question than trying to determine what things were. We try to focus on the latter consideration even if we also address the naturalness and fine-tuning issues.\\

Although it is in itself a well known solution to the Einstein field equations in cosmology, it is in practice often forgotten than the Big Bang singularity does not appear anymore when taking into account the possible positive curvature of the Universe mixed with an effective cosmological constant. It is regularly argued -- especially within the community of quantum gravity -- that new physics is unavoidably required to avoid the dramatic breakdown of general relativity (unless assuming very exotic contents) when going backward in time. This is not true when including the spatial curvature currently favored by observations. Several versions of this idea have already been discussed, for example in \cite{Martin:2003bp,Uzan:2003nk}. Interestingly, we will also show that this leads to original constraints of the reheating temperature and fixes the number of inflationary e-folds. Some considerations on the inflaton potential can be derived when facing the fine-tuning issue. This, finally, opens unique observational features through modifications of the primordial power spectrum.

\section{The curvature of the Universe}

In full generality (and in Planck units), the Friedmann equation reads
\begin{equation}
H^2=\frac{8\pi}{3}\rho+\frac{\Lambda}{3}-\frac{K}{a^2},
\end{equation}
where $H$ is the Hubble parameter, $\Lambda$ is the cosmological constant, and $K$ is the curvature. When one chooses $K=-1,0,+1$ this implies that the scale factor $a$ has dimension length. To remain consistent with the usual dimensionless choice for $a$, we assign to $K$ the dimension (length)$^{-2}$: it then measures the spatial curvature scale $^3R_0$. 
Recent results from the Planck collaboration \cite{Aghanim:2018eyx} slightly suggest a closed Universe: $-0.011>\Omega_{K,0}>-0.078$, using TT, TE, EE+lowE data, at the 95\% confidence level ($\Omega_K\equiv -K/(a^2H^2)$ and the subscript ``0" denotes quantities evaluated in the contemporary Universe).\\

Those conclusions where reanalyzed in details in \cite{Handley:2019tkm},  \cite{Park:2017xbl} and \cite{DiValentino:2019qzk}, strengthening the evidence for a positive curvature. Strikingly, it was shown that a positive curvature elegantly explains the anomalous lensing amplitude and removes the tension between the values of the cosmological parameters evaluated at different scales. This raises some discordances with baryonic acoustic oscillations, possibly revealing a crisis in cosmology. Still, it is concluded in \cite{DiValentino:2019qzk} that Planck data do favor a closed universe with a probability of nearly 99.99\%. In this work, we assume, as a hypothesis, that this result is correct. Although counter-arguments where given in \cite{Efstathiou:2020wem}, $\Omega_K>0$ is unquestionably consistant with, if not truly favored by, current observations. In addition, quite a lot of theoretical constructions, especially in the quantum gravity framework, do prefer a closed Universe. At this stage, it is fair to conclude that there is no conclusive proof of a positive curvature but this is a least an appealing possibility compatible with data.\\

A trivial -- but not so well-known -- feature of positively curved spaces is that they naturally bounce when the content is a positive cosmological constant. One can easily show that the Friedmann and Raychaudhuri equations admit an hyperbolic cosine solution:
\begin{equation}
a=\sqrt{\frac{3K}{\Lambda_i}}{\rm cosh}\left(\sqrt{\frac{\Lambda_i}{3}}t\right),
\end{equation}
where $\Lambda_i=8\pi\rho_{vac}$, with $\rho_{vac}$ the vacuum-like density leading to inflation. Among the large number of bouncing models \cite{Battefeld:2014uga,Lilley:2015ksa,Brandenberger:2016vhg}, this one has the appealing feature of not requiring any exotic physics. Pure general relativity (GR) with a positive curvature and cosmological constant does bounce and the past history of the universe is not singular, at least up to a hypothetic second bounce.\\

We therefore stress the following point: if we take the known laws of physics, start from the preferred values of the contemporary cosmological parameters and impose that an inflationary stage occurred in the past, we end up naturally with a bouncing ``origin" of the Universe. It is not useful to recall in this article the countless arguments in favor of inflation (the interested reader can, {\it e.g.}, go through \cite{Mukhanov:2005sc}). We assume here that inflation indeed took place in the early universe. The standard cosmological model, when evolved backward in time, then leads to a bounce. This quite trivial feature is not yet often considered seriously. 

It is however fair to also underline that other scenarii are hopefully  being considered: strictly speaking inflation is not mandatory. Several consistent alternatives to inflation can be found in \cite{Durrer:1995mz,Hollands:2002yb,Veneziano:2003enc,Brandenberger:2009jq,Creminelli:2010ba,Poplawski:2010kb,Lilley:2015ksa} and references therein.\\

The well-known singularity theorem of eternal inflation \cite{Borde:1993xh}, which generalizes the one of Hawking and Penrose \cite{Hawking:1966wvn}, holds strictly for open universes. A positive curvature anyway challenges eternal inflation \cite{Guth:2012ww}. Still, it should be mentioned that the probability of having no classical fluctuations at the end of the inflationary phase is very small. These fluctuations would become more and more important when going backward in time. This is the main argument of the Borde-Guth-Vilenkin singularity argument \cite{Borde:2001nh} and this would make spatial curvature less and less important as the universe contracts. 

\section{Duration of inflation and reheating temperature}

The contemporary universe can be thought of as made of four different cosmic ``fluids": the cosmological constant, matter, curvature, and radiation with respective current densities (ordered by decreasing value) $\rho_{\Lambda,0}$, $\rho_{M,0}$, $\rho_{K,0}$, and $\rho_{R,0}$. The Friedmann equation then simply reads $H^2=\frac{8\pi}{3}\sum_i \rho_i$, with a minus sign in front of the curvature term when one deals with a positive curvature. When going backward in time, those components grow at different rates: 
\begin{equation}
\rho_{\Lambda}\propto a^{0},\rho_{M}\propto a^{-3}, \rho_{K}\propto a^{-2}, \rho_{R} \propto a^{-4}.
\end{equation}
The density of radiation at the end of inflation is 
\begin{equation}
\rho_{R,RH}\approx\rho_{R,0}(1+z_{eq})^4\left(\frac{T_{RH}}{T_{eq}}\right)^4,
\end{equation}
where $z_{eq}$ and $T_{eq}$ are respectively the redshift and temperature at the time of equilibrium between matter and radiation and $T_{RH}$ is the temperature at the reheating that we assume to be sudden for simplicity\footnote{This is obviously a crude hypothesis, as demonstrated, {\it e.g.}, in \cite{Garcia:2020eof}.}. At that time, the radiation density was dominant and $\rho_{tot,RH}\approx\rho_{R,RH}$. The curvature density, although larger than the cosmological constant, was still smaller than the matter density and given by:
\begin{equation}
\rho_{K,RH}\approx\rho_{K,0}(1+z_{eq})^2\left(\frac{T_{RH}}{T_{eq}}\right)^2.
\end{equation}
When going further backward in the past, that is during the inflationary quasi-de Sitter stage, matter and radiation no longer exist anymore and the density of the scalar field (or whatever plays this role) remains constant at $\rho_{vac}\approx \rho_{R,RH}$. But the curvature density continues to grow.\\

Let us call $t_B$ the time when $\rho_{vac}=\rho_K$. This defines the precise moment when the Hubble parameter vanishes, that is the bouncing time -- this inevitably occurs. If there were $N$ inflationary e-folds between the bounce and the reheating (we assume here the transition between the bounce and the inflationary stage to be sudden too as it can easily be numerically shown that the associated number of e-folds is small), the curvature density at the bounce was $\rho_{K,B}=\rho_{K,RH}e^{2N}$. Setting $\rho_{K,B}=\rho_{vac}$ and using the value of $\rho_{K,0}$ favored by data \cite{DiValentino:2019qzk} leads to:

\begin{equation}
N\approx\frac{1}{2} {\rm ln} \left( \frac{\rho_{R,0}}{\rho_{K,0}} \left[ (1+z_{eq}) \frac{T_{RH}}{T_{eq}} \right]^2  \right).
\label{n}
\end{equation}

For $T_{RH}\approx 10^{16}$ GeV, this means $N\approx 65$.\\

This is an interesting result because this number is of the order of the smallest possible number of inflationary e-folds required to account for observational data. In the usual cosmological approach, this number can be anything above 65. Seen from this perspective, this approches sets a lower limit on the reheating temperature around the grand unification (GUT) scale: should $T_{RH}$ be smaller than this value, the inflation could not, in this model, have last long enough to solve the usual cosmological puzzles. This is 15 orders of magnitude above the experimental lower limit and quite close to the Planck scale. If this scheme of thought is correct, this basically fixes the number of inflationnary e-folds at $N\sim 70$, as Eq. (\ref{n}) leads, at most, to $N\approx72$ for $T_{RH}\approx T_{Pl}$ (which is anyway incompatible with observations). If $\rho_{K,0}$ is smaller than the value inferred from current CMB data, the number of e-folds will grow but the dependance is extremely weak and a variation of one order of magnitude on $\rho_{K,0}$ will change the number of e-folds by approximately $\Delta N \approx 1$. This basically means that whatever the measurable positive curvature of the Universe, our conclusions do hold, in agreement with \cite{Uzan:2003nk}.\\

In principle, this statement could be relaxed. If one simply requires the number of inflationary e-folds to be equal to the number of post-inflationnary e-folds, without requiring $N>65$, then any curvature density currently comparable to the radiation density would work and a low-scale inflation scenario could be conceivable. This, however, conflicts with data for reasons that will be made explicit in the last section. 

\section{Fine-tuning and inflation potential}

This study suggests that, taking into account the known laws of physics, the measured contemporary cosmological parameters (including curvature), and forcing the history of the Universe to go through ``events" that we highly suspect to have occurred (including inflation), a bounce should have taken place and the reheating temperature should be around the GUT scale. This way of reconstructing the past is the one used by archeologists or historians and is perfectly legitimate. What this kind of thinking however does {\it not} say is ``why"  this has occurred.\\

If the sequence is now considered with time flowing in the usual direction, this scenario clearly raises questions. It is well known that a massive scalar field will generally not be, during the contraction phase, in the strongly potential energy dominated regime required for the bounce to occur. For the curvature term to dominate the dynamics and induce the bounce, the equation of state parameter, $w=p/\rho$, has to remain durably smaller than -1/3. This is a possible solution of the equations of motion but this is not at all a dynamically favored situation. Basically, one has to ``choose" a trajectory with $w\approx -1$.\\

The latter statement is however intricate. Obviously, when an astronomer deals with the collision of comet and a planet, she does not care about the {\it a priori} low probability of this event. She just studies what has happened and how it can be used to understand better the World and its properties. We are somehow in a comparable situation. Every single trajectory is, by definition, of zero mesure in the continuous parameter space \cite{Page:1984qt}. The construction of a meaningful bayesian estimator is a hard task that highly depends on the chosen priors. For exemple, even within the narrow community of loop quantum cosmology, there is a lively debate on the ``most probable" number of inflationary e-folds predicted by the model after the bounce. The approach advocated in \cite{Ashtekar:2009mm,Ashtekar:2011rm} does not agree at all with the one pushed in \cite{bl,Linsefors:2014tna,Bolliet:2017czc,Martineau:2017sti}. Another known huge discrepancy is between the probability for inflation estimated in \cite{Gibbons:2006pa} and the one calculated in \cite{Linde:2007fr}. The conclusions are clearly in strong opposition and it is safe to conclude that evaluating the probability of a single situation is extremely difficult. General explanations are given in \cite{Schiffrin:2012zf}. This is, by the way, less due to the so-called ``measure problem in cosmology" than to the trivial fact that finding a dynamical variable to which a known probability distribution function (PDF) can be assigned is extremely difficult.\\

Evaluating the ``naturalness" of this trajectory is beyond the scope of this study and would require the existence of an uncontroversial mesure or PDF, which is anyway missing. This is, in many senses, an ill-defined question. We prefer, here, to stress what the past was -- knowing what we know -- and not why it was so.\\

If the issue of fine-tuning is however taken seriously, one should consider two different ways of thinking. The first one is the grounded in the Multiverse framework. In such a view, the anthropic bias should be taken into account and it could very well be that contracting branches where the field does not lead to a bounce simply die, without developing observers. We therefore naturally find ourselves on a very specific trajectory compatible with a bounce in the past, as we would otherwise not exist. The subtle question of the emergence of observers in the contracting branch is obviously beyond our curent knowledge. It is therefore very hard at this stage to make clear predictions in this framework \cite{Carr:2007zzb}.\\

It could also be argued that any amount of radiation or cold matter in the contracting phase would prevent the spatial curvature from being important at high densities. This is correct unless a deflation stage occurs, which is anyway, in itself, required by the model.\\

More interestingly, if one focuses on a single universe and chooses a simple measure, {\it e.g.} based on the phase of the scalar field in the contradicting blanch, then flat potentials are favored  \cite{Matsui:2019ygj}. In particular, potentials of the form
\begin{align}
V(\phi) =& V_0 \left( \tanh ^{ 2 }{ \left[ \frac { \phi  }{ \sqrt { 6\alpha  }  }  \right]  } +\beta \tanh { \left[ \frac { \phi  }{ \sqrt { 6\alpha  }  }  \right]  } +\gamma  \right), \label{V}
\end{align}
with $\alpha > 0$, $-1 < \beta < 1$, and $-1 < \gamma \leq 0$ do lead to the desired equation of state of the scalar field during contraction.  Importantly,  this could be compared with CMB measurements as the shape of the potential begins to be quite well constrained \cite{Ade:2015lrj}. But this also opens a new indirect way to determine acceptable potential shapes.

\section{Consistency}

First, it is worth noticing that many curvature bounces might have occurred in the past, that is before the bounce which gave rise to our expanding branch. Basically, a bounce takes place each time the curvature density becomes equal to the sum of the energy densities of all the other fluids in the Universe. In the expanding branch, the curvature density inevitably dominates at some point, unless the cosmological constant begins to dominate before this happens. This is precisely the case in our universe and this is why we are ``protected" from a curvature bounce that would otherwise have happened in a quite near future. When going backward in time, before the bounce, the same situation happens. Either we are on a trajectory with a ``long enough" deflation stage (a certain amount of deflation is anyway required for the bounce to happen) and the cosmological constant might make the bounce unique. Or the deflation was quite brief -- which is favored by naive measures -- and another crossing took place in the past (unless the universe was filled with an exotic content described by an unusual equation of state). This does not lead to any phenomenological difficulty. This, however, revives the singularity issue.\\

As in all bouncing models one might raise entropic concerns. Due to the complexity of the definition of the gravitational entropy, this important consideration is highly non-clear and should probably be addressed from a ``relational" point of view \cite{Rovelli:2018vvy}. It even makes sense to question seriously the direction of time in preceding the branch: in the ``oriented coarse graining" hypothesis \cite{Rovelli:2014cja}, the thermodynamical time might flow in the opposite direction.\\

A more important and obvious concern is about quantum effects. One might argue that even if a purely classical bounce -- as the natural consequence of current observations -- is appealing, it could remain problematic that quantum gravity is ignored. Although speculative and still under construction \cite{Oriti:2009zz}, quantum gravity is anyway expected to play a crucial role in the early universe. Avoiding the use of quantum gravity is tantalizing, but how reliable would the predictions then be ? The remarquable point that we make here is that when going backward in time from what we know, the Universe never approaches the Planck density. The bounce occurs at the inflationary scale, which is experimentally known to be much smaller than the Planck scale (thanks, {\it e.g.}, to the upper bound on the tensor-to-scalar ratio, $r<0.056$ \cite{Akrami:2018odb}). The model is therefore consistent and never drives the Universe into a ``quantum cosmology" stage. It also seems safe to ignore strong backreaction effects. Although there are many excellent reasons to try building a quantum theory of gravitation, the statement that quantum gravity is necessary for consistency reasons when studying the early universe might be wrong.\\

Still, it is important to underline that although inflation is part of the standard model of cosmology, it might itself somehow ``require" new physics. After all, the only known fundamental scalar field cannot be the inflaton (at least with usual couplings). In addition, the use of effective field theory methods beyond the realm where they can be applied is hazardous. There have recently been severe challenges to inflation and doubts raised on its consistency with fundamental physics (in particular by the ``swampland'' criteria \cite{Brown:2015iha,Brahma:2019iyy,Wang:2019eym}). We adopt here the view that either requiring exotic physics or not, inflation did happen (which is actually not mandatory -- see, {\it e.g.}, \cite{Peter:2009sa} and references therein for some possible ways to observe alternatives to inflation).\\
 
Finally, it is important to check that the number of e-folds
\begin{equation}
N=\int\frac{H}{\dot{\phi}}d\phi,
\end{equation}
calculated from the dynamics of the field in the quadratic potential\footnote{We leave for a future study the generalization to other potentials but the main idea remains quite generic.} is compatible with the number of e-folds required by the consistency of the model. It could, {\it a priori}, be that the equations of motion do not lead to a long enough inflationary stage for the proposal to be convincing. It is however easy to check that an initial value of the field around $\phi\sim16.7-17$ (depending on the precise value chosen for the curvature) does work and satisfies all the constraints. 

\section{Observational footprints}

The main observational consequences of a positive curvature are known. This leads to an infra-red suppression of the power spectrum, around the curvature scale \cite{Efstathiou:2003hk}. In particular, the abnormally low power measured for very small $\ell$'s might be explained by this effect, although the statistical significance is too low to lead to any firm conclusion. Recently, this has been readdressed in details \cite{Bonga:2016iuf}. One of the effects of the curvature is to make the wavenumbers of the quantum perturbations discrete, due to the spherical topology of the spatial sections. This plays a role in the post-inflationary era. The other main effect is obviously due to the modification of the Friedmann equations which plays a role at the beginning of inflation. \\

However, the calculations of \cite{Bonga:2016iuf,Bonga:2016cje} were made within a pure inflationary framework without any bounce. If the scenario presented in this article is correct, it raises a very exciting possibility for observations. The main problem with experimental footprints of bouncing models including a stage of inflation is that inflation generically lasts so long that the portion of the primordial power spectrum which is probed by observations is too far in the ultra-violet to be probed \cite{Barrau:2017rwl}. Otherwise stated, the interesting distorsions of the spectrum induced by the bounce are usually not observable in the CMB because inflation has ``thrown them away" far beyond the horizon.\\

The fact that the number of inflationary e-folds is, in the approach advocated here, constrained to around 70 has the very interesting consequence that it makes the effects of the bounce potentially observable. The system to be solved is well defined:

\begin{eqnarray}
\dot{H} &=& - 4 \pi \left( \rho + p \right) + \frac{K}{a^2},
\label{eq:friedmann} 
\\
 \ddot{\phi}  &=& - 3 H \dot{\phi} - \frac{dV}{d\phi} ,
\label{kg}
\\
\ddot{q} &=&  b(n, t) \dot{q} + c(n, t) q,
\label{pot}
\end{eqnarray}
where the two first equations define the background evolution (the Raychaudhuri equation is usually numerically more tractable than the Friedman one) and the last one refers to gauge invariant perturbations $q$ in hyperspherical topology, $b(n, t)$ and $c(n, t)$ being functions of the wavenumber label $n$ and of the inflaton potential. The explicit calculation will be performed elsewhere \cite{prep}. It is however clear that the limited number of inflationary e-folds will make footprints of the bounce potentially visible. At variance with the simple power suppression associated with the curvature, the bounce will produce oscillations in the spectrum that could, in principle, be disentangled. The usual consistency relation of inflation will be violated.\\

This also raises an important point. Most of the usual cosmological puzzles can be solved by requiring the number of inflationary e-folds to be at least equal to the number of post-inflationary e-folds. Although quite unusual it is, in some cases, possible to consider scenarios with $N<65$. This is why the argument given at the beginning of this work to set the lower limit on $T_{RH}$ might seem weak: if one chooses a lower value of $T_{RH}$, both the number of e-folds of inflation and after inflation will be decreased by the same amout and the model will remain {\it a priori} correct. However, as the ``non-trivial" effects on the power spectrum appear for comobile numbers of the order of one (and less), a low value of $T_{RH}$, would inevitably shift the observational window to the infrared part of the spectrum which is {\it not} scale-invariant. As $a_{CMB}\sim e^{N}T_{RH}/T_{CMB}$, for $a_{B}=1$, lowering $T_{RH}$ (and, necessarily, $N$) will shift the physical wavenumber\footnote{Rigorously speaking, instead of expanding the equation of motion in the Fourier space, one needs to expand the perturbations on the tensor hyperspherical harmonics and the wavenumber mentioned here is ``effective" and discrete.} to the portion of the spectrum affected by the bounce. This is would be inconsistent with data and this is why the model basically fixes the duration of inflation and the reheating temperature. Although we leave the accurate calculation for a future study, the basic argument is simple and fixes the orders of magnitude.\\

Another quite specific feature of this model is the possibility to ``see through" the bounce. Beyond indices that could be seen in the CMB, it might also be possible to detect gravitational waves from events taking place in the contracting branch. This is plausible because of the low energy scale at which the bounce occurs in this approach. The very weak coupling of gravitational waves allow them to cross the bounce when the density remains much small than the Planck density. But, more importantly, this possibility might be realistic because of a subtle behavior of the luminosity distance \cite{Barrau:2017ukm}. In the contracting phase, between an event at (negative) time $t_e$ and the detection of the associated signal at (negative) time $t_r$, the luminosity distance reads

\begin{equation}
D_L=c\frac{(-t_r)^{2n}}{n-1}\left[ \frac{(-t_r)^{1-n}}{(-t_e)^n}-(-t_e)^{1-2n} \right],
\label{DL5}
\end{equation}
where $n$ is defined by the scale factor evolution: $a(t)=k(-t)^n$. For a dust-like content, the luminosity distance does {\it grow} with cosmic time. Although counter-intuitive at first sight, this behavior is due to the amplification associated with the contraction which counter-balance the dilution of the propagation. For a deflation stage, it reads 
\begin{equation}
D_L=c\frac{e^{\alpha (t_e -2t_r)}}{\alpha}\left[ e^{\alpha t_r}-e^{\alpha t_e} \right],
\label{DL6}
\end{equation}
where $\alpha=|H|$ is such that $a(t)=ke^{-\alpha t}$. Those considerations have, up to know, been only considered for inflationless models (in the expanding branch following the bounce) as the inflationary stage would dilute the signal. However, in the framework considered here the unavoidable deflation stage changes the game. If the duration of deflation was comparable to the one of inflation, it is easy to check that events like the coalescence of massive black holes in the contracting branch could  still be measured today.  

\section{Conclusion}

In this article, we have shown that evolving backward in time the currently observed state of the Universe, and imposing a hard-to-avoid inflationary stage, leads to a classical bounce without the need for speculative physics. No new ingredient is required. The result depends crucially on the existence of the now preferred positive curvature of the Universe but not on its precise value. This trajectory might be ``unstable" but it remains possible and could be observationally favored. In addition, the instability is basically due to the description of inflation by a scalar field which is not the only way to produce a vacuum-energy dominate stage. We also show that this somehow ``fixes" the number of inflationary e-folds, the inflation energy scale, and opens exciting possibilities for observational tests, while being consistant with a purely classical treatment. The evolution is always classical and deterministic.\\

Many questions however remain open at this stage. We have considered here a purely isotropic case. This is not realistic as anisotropies inevitable grow during the contraction phase. The subtle effects of the cosmic shear should be accounted for.\\

Another important and interesting issue is the one of initial conditions for perturbations. This has been disregarded when stating that the system to solve is well defined. A key problem, especially for scalar perturbations known for being associated with intricate potentials, is to define a natural initial state when a clear Bunch-Davies like vacuum does not exist anymore. This is a substantial difficulty for bouncing models \cite{Schander:2015eja} which has to be treated here.

\bibliography{refs}

 \end{document}